# On Using Magnetic and optical methods to determine the size and characteristics of nanoparticles embedded in oxide semiconductors

Gillian A. Gehring, Harry J. Blythe, Qi Feng, David S. Score, Abbas Mokhtari, Marzook Alshammari, Mohammed S. Al Qahtani and A. Mark Fox
Department of Physics and Astronomy, University of Sheffield, Sheffield S3 7RH, United Kingdom.

**Films of oxides doped with transition metals are frequently believed to have magnetic inclusions. Magnetic methods to determine the amount of nanophases and their magnetic characteristics are described. The amount of the sample that is paramagnetic may also be measured. Optical methods are described and shown to be very powerful to determine which defects are also magnetic.**

## I. Introduction

Oxide semiconductors for example, ZnO, $In_2O_3$, $TiO_2$ and $SnO_2$ are frequently ferromagnetic at room temperature if grown including a small concentration of a transition metal (TM) [1]-[3] which means that they are possible candidate materials for the next generation of spintronics devices [4]. This has led to a major research effort devoted to TM doped oxides, but the mechanism that causes the ferromagnetism remains controversial, and its usefulness is limited by the small values of both the coercive field and remanence [1]. Since magnetism is observed for low concentrations of dopants and measurements of XMCD have not detected magnetism of the core states of the transition metal ions [5], there has been much effort to understand the mechanism. In some cases it is clear that the observed magnetization comes from blocked nanoparticles of an impurity phase such as metallic cobalt or iron or of an impurity oxide phase [6]. It is important to establish the extent to which this is a bulk phenomenon in which case there should be polarized host electrons that can be used for spintronics applications and to whether it arises solely because of ferromagnetic inclusions

Various physical methods have been used to observe these inclusions. These include structural studies, such as HRXRD, TEM and EXAFS [7,8], to look for small inclusions and the structural coordination around the TM ion. X-ray spectroscopy detects the valence state of the metallic element. This can indicate the presence of the metallic phase [9] or a magnetic oxide phase e.g. $Fe_2O_3$ or $Fe_3O_4$ in the case of an Fe dopant [2].

These structural methods do not specify the magnetic state of the nanoparticle. In some cases it has been established that the defect region that is observed in a structural study may have a core which is metallic but surrounded by an amorphous region [10]. If the surface of the nanoparticle is rough a sizeable fraction of the atoms may have their moments coupled antiferromagnetically through neighboring oxygen ions.

Since we are primarily interested in the magnetic effects of any nano-clusters it is sensible to investigate the range of magnetic probes that can be used. In this paper we concentrate on the magnetic and magneto-optical methods available to determine the fraction of the magnetization that may be present in the nanoparticles, the fraction of the TM ions that are paramagnetic down to low temperatures and the magnetic anisotropy of the nanoparticles. The methods are known but it is useful to summarize them here and apply them to thin films of doped oxides. We present data for a sample of $Zn_{1-x-y}Co_xAl_yO$ with x=0.05 and y=0.006, which was deposited by pulsed laser deposition using a Lambda Physik XeCl laser (λ=308 nm) at a repetition rate of 10 Hz; we determined that a small percentage of the Co is metallic [9]. All the data presented here except that from Fig. 1, which is from a sample of doped $In_2O_3$, is from this sample.

## II. The Paramagnetic Fraction

The total magnetization of thin films typically a few 100 nm thick and always < 1μm is dominated by the diamagnetic contribution of the substrate which is ~0.5mm thick. A typical plot for a weakly magnetic sample is shown in Fig 1. This sample was Fe doped $In_2O_3$ grown at 100 mTorr [11], chosen because it shows this effect clearly. The linear part at high fields, usually classified as the diamagnetic contribution of the substrate is temperature dependent and this is due to a paramagnetic or superparamagnetic contribution from the film at low temperature. The true diamagnetic susceptibility of the substrate will be independent of temperature and so it is easy to check if there is a paramagnetic contribution from any temperature dependence of the linear term. A more accurate approach is to measure a substrate directly and measure the deviation between the subtracted term and the true diamagnetic susceptibility scaled for the area of the sample, this agrees with that measured at room temperature for this sample. This is a clear case of a sample that shows that a paramagnetic contribution from the film increasing at low temperatures.

## III. Magnetism from Superparamagnetic Inclusions

Above their blocking temperature $N$ superparamagnetic particles with individual magnetizations $m_{eff}$ are in equilibrium with the lattice. The total magnetization, $M$, follows a Langevin function,

$$M(H,T) = Nm_{eff}\left[\coth x - \frac{1}{x}\right]; \qquad x = \frac{m_{eff}\mu_0 H}{k_B T}. \quad (1)$$

In the low field limit this expression reduces to Curie's law,

$$M(H,T) = \frac{Nm_{eff}^2 \mu_0 H}{3k_B T}. \quad (2)$$

Below the blocking temperature the particles are not able to reorient fast enough to stay in equilibrium and so the magnetization becomes dependent on its history. This is seen in field cooled and zero field cooled scans. The zero field cooled sample is expected to show a peak in the magnetization at the blocking temperature [6] where both the coercive field and the remanence fall to zero. However this is not always seen. Fig. 2 shows plots of the coercive field and the remanence for doped ZnO – it is seen that they fall rapidly but do not actually go to zero. We show here that we can determine the contribution to the magnetization from blocked particles even when we do not have a peak in the zero field cooled magnetization.

We use another useful relation for blocked particles which is that is that the ratio of the remanence to the saturation magnetization, $M_r/M_s$ is equal to ½ if they are oriented at random. We can use this relation and the value of the saturation magnetization $M_s = Nm_{eff}$ fit to the Curie law given above in equation (2). At a given temperature $T$ the saturation magnetization of the particles that have become unblocked is given by,

$$M_{unblockedclusters}(T) = 2\left(M_r(0) - M_r(T)\right) = Nm_{eff} \quad (3)$$

We estimate $m_{eff}$ from the blocking temperature ( taken to be 250K) assuming that the clusters of metallic cobalt occupy a volume $V$ using $m_{eff} = Vm_{Co}$ and $KV = 30k_B T_B$ where $K$ is the anisotropy. Above the blocking temperature the low field susceptibility is measured directly and compared with the Curie law as calculated as above,

$$\chi(T) = \frac{2\left(M_r(0) - M_r(T)\right) m_{eff}\mu_0}{3k_B T}. \quad (4)$$

The results are shown in Fig. 3. Although the errors are large in the calculated values because of the uncertainty in estimating $m_{eff}$, there is qualitative agreement indicating that the superparamagnetic particles of metallic cobalt contribute about 2/3 of the magnetization in this case. This indicates that it is possible to get a reliable estimate of the contribution from blocked particles even when the blocking temperature is high so that there is still a coercive field at room temperature.

## IV. Contribution to the Magneto-Optics from Impurity Phases

Magneto-optics is an ideal tool to look for impurity phases of insulating compounds if their band gap is less than the host as is often the case for ZnO and $In_2O_3$ which have wide band gaps~3.5eV. For example, oxides of Mn have band gaps around 2-2.5eV and a magneto-optic signal was detected from the impurity band edge at low temperatures in Mn doped ZnO [12].

It is also straightforward to study metallic inclusions by magnetic means as the Maxwell–Garnett theory is well established and the dielectric tensor of metallic cobalt is well known [10], [13], [14]. It is found that the Maxwell-Garnett theory fits the data well for wide band gap materials such as MgO ($E_{gap}$~7eV)[10] provided that the electron scattering time, τ, in the cobalt clusters is reduced from the bulk value. However the theory predicts that the energy for which Im $\varepsilon_{xy}^{eff}(\omega) = 0$ is independent of the value of this relaxation time, τ, and depends only on the demagnetization factor of the cobalt clusters, $L_x$, the refractive index of ZnO, $n$, the fraction of the whole material that is metallic cobalt, $f$, and the plasma frequency of cobalt, $\omega_p$. The frequency of the crossing point is given by, $\omega_X^2 = \frac{a}{n^2(1-a)+a}\omega_p^2$ where $a = L_x(1-f)$ [13]. A plot of the variation of the crossing point with $L_x(1-f)$ is given in Fig. 4.

The frequency dependence of the imaginary part of the dielectric constant for this sample of ZnO with dissolved cobalt is given in Fig. 5 and compared with the Maxwell-Garnett theory. In this case we see that there is partial agreement with the Maxwell-Garnett theory but that there is also a contribution that is arising from Co induced magnetism in the ZnO lattice which results in the larger than expected energy at which Im $\varepsilon_{xy} = 0$ . This would result from a negative contribution to Im$\varepsilon_{xy}$ at the ZnO band gap energy of ~3.4eV. There is also a noticeable feature at lower energy. The weak dispersive feature at ~2eV is indicative of $Co^{2+}$ ions contributing to the magnetism in the sample. Hence this data shows us that the Co nanoparticles are not the only source of magnetism in this material.

## V. Conclusions

We see that the contribution to the magnetism from isolated ions and nanoclusters may be ascertained from both magnetic and optical means. Together this enables us to determine the existence of nano-clusters and also to ascertain if they are the only source of magnetization in these materials.

## VI. Acknowledgments

We would like to thank EPSRC for funding for equipment and also studentships for D. S. Score and K. Addison and K.A.C.S.T for funding for studentships for M. Alshammari and M. S. Al Qahtani. We are grateful to Xiao-li Li, Zhi-yong Quan and Xiao-Hong Xu for letting us include data on their sample of doped $In_2O_3$.

## References

[1] J. M. D. Coey "High-temperature ferromagnetism in dilute magnetic oxides" *J. Appl. Phys.* vol. 97, pp. 10D313, 2005.

[2] X.H. Xu, FX Jiang, J Zhang, XC Fan, HS Wu and GA Gehring **"**Magnetic and transport properties of n-type Fe-doped $In_2O_3$ ferromagnetic thin films" *Appl Phys Lett* vol. 94, pp. 212510 , 2009.

[3] J.D Bryan et al., *J. Am. Chem. Soc*. vol. 126 (37), pp. 11640, 2004

[4] S.J. Pearton et al., *Semicond Science and Tech* vol. 19, pp. R59, 2004.

[5] T. Tietze et al., "XMCD studies on Co and Li doped ZnO magnetic semiconductors" *New Journal of Physics* vol. 10, pp. 055009, 2008.

[6] M. Opel et al., "Nanosized superparamagnetic precipitates in cobalt-doped ZnO" *Eur. Phys. J. B* vol. 63, pp. 437, 2008.

[7] Shengqing Zhou et al., "Crystallographically oriented Co and Ni nanocrystals inside ZnO formed by ion implantation and post annealing" *Phys Rev B* vol. 77, pp. 035209, 2008.

[8] T.C. Kaspar, T. Droubay, S.M. Heald, M.H. Engelhard, P. Nachimuthu, and S.A. Chambers "Hidden Ferromagnetic Secondary Phases in Cobalt-doped ZnO Epitaxial Thin Films." *Phys Rev. B* vol. 77(20), pp. 201303, 2008.

[9] S.M. Heald, T. Kaspar, T. Droubay, V. Shutthanandan, S. Chambers, A. Mokhtari, A.J. Behan, H.J. Blythe, J.R. Neal, A.M. Fox and G.A. Gehring "X-ray absorption fine structure and magnetization characterization of the metallic Co component in Co-doped ZnO thin films" *Phys Rev B* vol. 79, pp. 075202, 2009.

[10] C. Clavero, A. Cebollada, G. Armelles, Y. Huttel, J. Arbiol, F. Peiró and A. Cornet *Phys Rev B* vol. 72, pp. 02441, 2005

[11] Xiao-Hong Xu, Feng-Xian Jiang, Jun Zhang, Xiao-Chen Fan, Hai-Shun Wu and G. A. Gehring, "Magnetic and transport properties of *n*-type Fe-doped $In_2O_3$ ferromagnetic thin films" *Appl. Phys. Lett.* **94**, 212510 (2009)

[12] J.R. Neal , A.J. Behan, R.M. Ibrahim , H.J. Blythe ,M. Ziese, A.M. Fox and G. A. Gehring *Phys. Rev. Lett.* vol. 9**6,** pp. 197208, 2006.

[13] David S Score, Marzook Alshammari, Qi Feng , Harry J. Blythe, A. Mark Fox, Gillian A. Gehring, Zhi-Yong Quan, Xiao-Li Li and Xiao-Hong Xu "Magneto-optical properties of Co/ZnO multilayer films" *Proc ICM to appear and Cond Matt* 0909.4149, 2009

[14] G. S. Krinchik . *J. Appl. Phys.* vol. 35, pp. 1089, 1964.

## Figure Captions

Fig. 1 (color on line) SQUID measurement of the magnetization of $In_2O_3$ doped with 5% of Fe and grown at 100mTorr on a sapphire substrate showing the paramagnetic contribution to the 'diamagnetic background' at low temperature.

Fig. 2 (a) the coercive field, $H_c$, and 2(b) the magnetization remanence divided by the magnetization saturation, $M_r/M_s$, as a function of temperature.

Fig. 3 (color on line) χ(T) as calculated from equation (4) using the measured $M_r$ and $m_{eff}$ deduced from blocking temperature compared with the measured χ(T) from the hysteresis loops.

Fig. 4. A plot of the crossing frequency expected for the MCD in terms of $L_x$ and *f*. For a small fraction of metallic cobalt, *f*, as we have here we expect that the clusters will be spherical on average so $L_x$ ~1/3 and hence $\omega_x$~2.8eV.

Fig. 5 (color on line) The dielectric function Im $\varepsilon_{xy}$ determined from MCD measurements and the effective dielectric function Im $\varepsilon_{xy}$ calculated according to Maxwell-Garnett theory for *f*=0.006, and different $L_x$ and *τ*. The crossing point is independent of the relaxation time, τ.

Figure 1

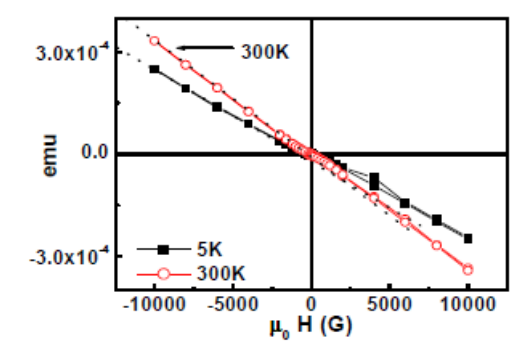

Figure 2

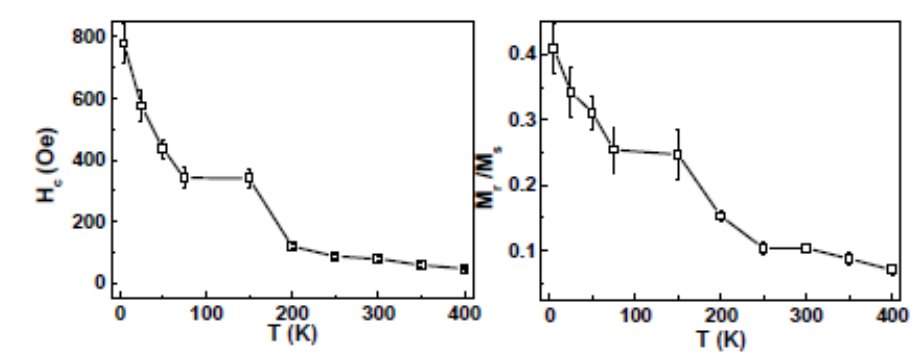

2(a) 2(b)

Figure 3

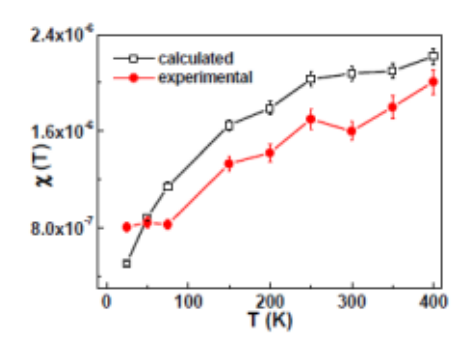

Figure 4

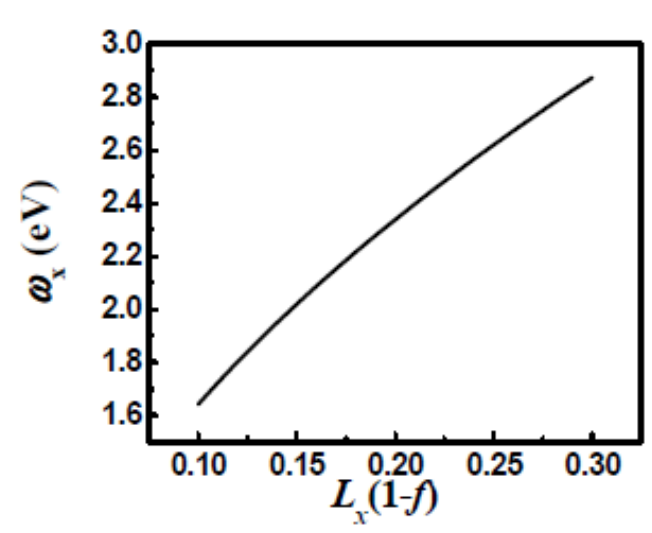

Figure 5

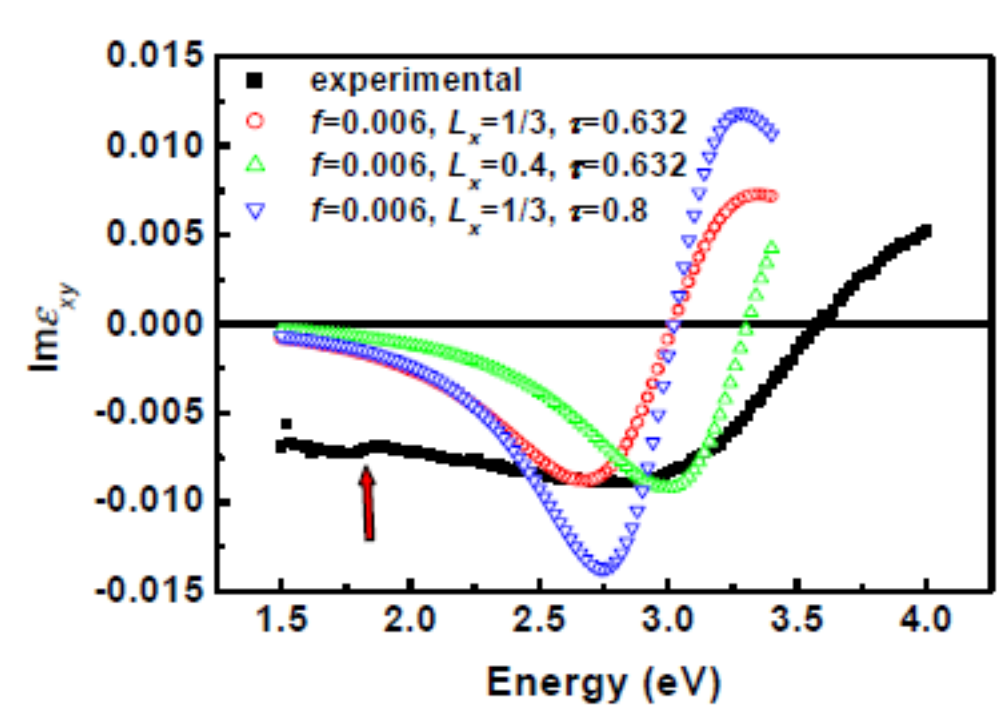